\documentclass[aps,prl,twocolumn,notitlepage,superscriptaddress,showpacs]{revtex4}
\usepackage{graphicx} 
\usepackage{amsmath}
\usepackage{amsfonts}
\usepackage{mathrsfs}
\usepackage{bm}
\usepackage{color}

\begin{document}
\title{Non-local quantum fluctuations and fermionic superfluidity in the 
imbalanced attractive Hubbard model}

\author{M.~O.~J.~Heikkinen}
\affiliation{COMP Centre of Excellence and Department of Applied Physics,
Aalto University, FI-00076 Aalto, Finland}
\author{D.-H.~Kim}
\affiliation{Department of Physics and Photon Science, Gwangju Institute of Science and Technology, Gwangju 500-712, Korea}
\author{M.~Troyer}
\affiliation{Theoretische Physik, ETH Zurich, 8093 Zurich, Switzerland}
\author{P.~T\"{o}rm\"{a}}
\email{paivi.torma@aalto.fi}
\affiliation{COMP Centre of Excellence and Department of Applied Physics,
Aalto University, FI-00076 Aalto, Finland}

\begin{abstract}
We study fermionic superfluidity in strongly anisotropic optical lattices
with attractive interactions utilizing 
the cluster DMFT method, and
focusing in particular on the role of non-local quantum fluctuations.
We show that non-local quantum fluctuations impact the BCS superfluid transition
dramatically. 
Moreover, we show that exotic superfluid states with delicate
order parameter structure, such as the Fulde-Ferrell-Larkin-Ovchinnikov phase
driven by spin population imbalance, can emerge even in the presence of
such strong fluctuations.
\end{abstract}

%71.10.Fd Lattice fermion models (Hubbard model, etc.)
%67.85.-d Ultracold gases, trapped gases (see also 03.75.-b Matter waves in quantum mechanics)
%03.75.Ss Degenerate Fermi gases

\maketitle

Mean-field theories have been tremendously successful at furthering our understanding
of quantum many-body physics. For instance, the explanation of conventional superconductivity
based on the BCS-theory is hailed as one of the highest achievements in condensed matter physics. 
Nonetheless, it is well-known that the mean-field treatment in general can facilitate
a qualitative description of the physical system -- at best. 
In the context of lattice models, dynamical mean-field theory (DMFT) 
constitutes a substantial improvement over static mean-field treatments by including
fully the effect of 
local
quantum fluctuations. Yet, even the predictions of DMFT 
may fail in the presence of \textit{non-local} quantum fluctuations, i.e.~non-local
contributions to the self-energy of the system.
Ultimately, the emergence of an ordered phase can be firmly predicted
only if the non-local quantum fluctuations are properly accounted for.
Moreover, key information of the physical system can be encoded to the
non-local structure of the self-energy. This is true for example for
the d-wave symmetry of high temperature superconductors.

The elusive, yet ubiquitous, nature of non-local quantum fluctuations raises the
question, whether it is possible to identify physical systems where the effects of these fluctuations 
could be studied in a systematic manner. 
In this respect, ultracold gas setups with controllable dimensionality 
seem to offer a natural path forward regarding that the non-local fluctuations 
are most prominent in low-dimensional systems.
The dimensional crossover from 1D to higher dimensional systems has garnered broad interest.
From the theoretical point of view, 
it is anticipated that phases of matter prominent in 1D models can be stabilized 
when brought to a higher dimensionality~\cite{GiamarchiBook}.
Experimentally, the strong dimensional anisotropy may also offer advantages over
a more straightforward 3D geometry, as demonstrated in a recent work on
repulsively interacting fermions in an anisotropic optical lattice,
where the temperature scale of anti-ferromagnetic correlations 
was reached~\cite{Greif2013,Sciolla2013,Imriska2013}.

One of the most intriguing many-body phenomena which can be approached in the
context of dimensionally tunable lattices is that of fermionic superfluidity.
The paradigm case of fermionic superfluidity with s-wave spin-singlet BCS pairing 
could be studied in an experimental realization of the
attractively interacting Fermi-Hubbard model~\cite{Bloch2008review}.
Moreover, there is a wide consensus that this system 
might demonstrate exotic forms of superfluid pairing 
when subjected to e.g.~a spin population imbalance. 
The prospects of realizing
such forms of conventional and exotic superfluidity in systems of intermediate dimensionality 
have been discussed broadly in the 
literature~\cite{Klemm1976,Schulz1983,parish2007,zhao2008,Kim2012,Heikkinen2013}.
However, the role of non-local quantum fluctuations remains to a large degree
an open question in these systems even in the case of the conventional BCS pairing.

\begin{figure}
\includegraphics[width=0.37\textwidth]{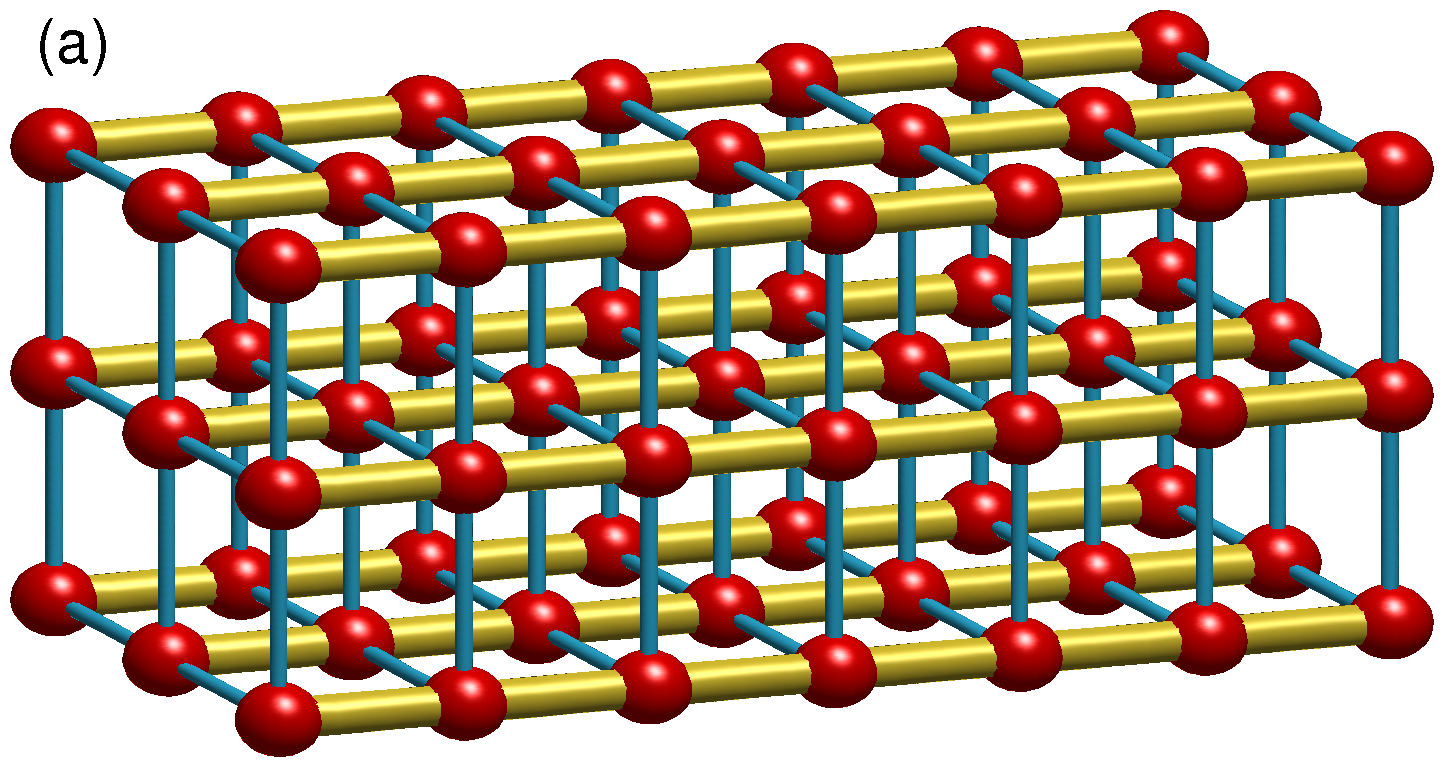}
\includegraphics[width=0.44\textwidth]{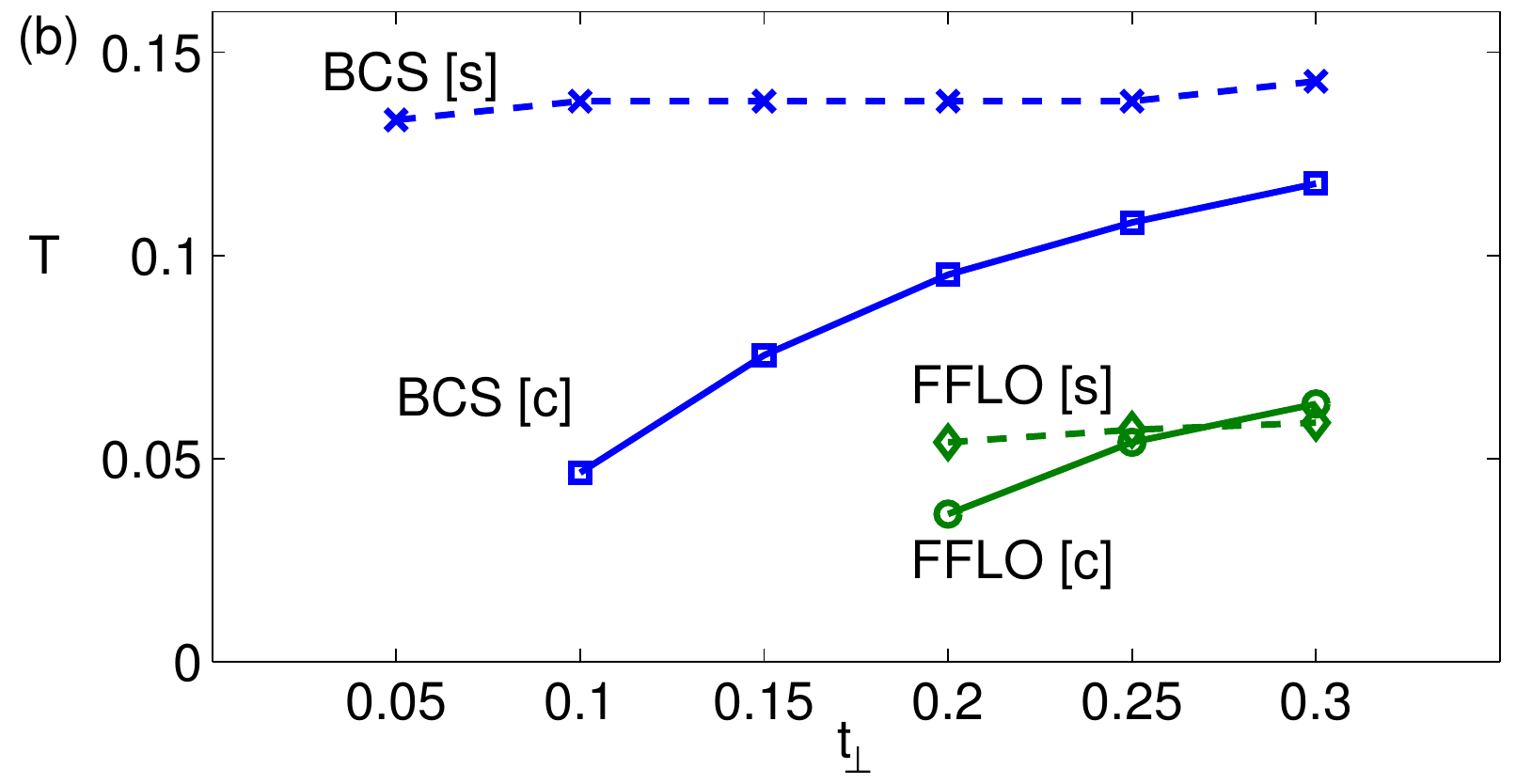}
\caption{\label{fig:phasediag}
(a) A schematic of the system geometry. The optical lattice consists
of one-dimensional chains which are coupled to form an anisotropic
cubic lattice. In our cluster DMFT scheme, the 1D chain
is treated as a single cluster with periodic boundaries.
(b) The BCS and FFLO phase transitions as a function of the interchain hopping $t_\perp$
both in the cluster DMFT model (with non-local quantum fluctuations)
and the real-space DMFT model (excluding non-local quantum fluctuations).
The critical temperature of the BCS state is given by the continuous
and dashed blue line for the cluster [c] and single-site [s] models, respectively.
Similarly, the FFLO critical temperature is given by the green line for the cluster [c] model
and by the dashed green line for the single-site [s] model.
As a point of contrast, the static mean-field prediction of the BCS critical temperature
is $T_{c,\textrm{MF}}= 0.52$ in the corresponding parameter range, 
while the FFLO critical temperature would be on the order of $0.5~T_{c,\textrm{MF}}$~\cite{Koponen2007}.
In each case the system is at half-filling, and the
FFLO transition is found by varying the spin-polarization
of the system while maintaining the total filling fraction constant.
For $t_\perp\geq 0.15$, we perform the calculations in a cluster of $N_c=36$ 
lattice sites, whereas for $t_\perp = 0.1$ a cluster size of $N_c=42$ is
required for convergence.}
\end{figure}

In this work, we study an attractively interacting two-component Fermi gas 
in a strongly anisotropic cubic optical lattice, 
see Fig.~\ref{fig:phasediag}(a). We compute the phase diagram of this
system using cluster and real-space variants of DMFT,
and investigate the effect of non-local quantum fluctuations on the different
possible forms of superfluidity occurring in the system.

The system is described by the Hubbard Hamiltonian
\begin{align}
\mathcal{H} = & -t_\parallel \sum_{jl\sigma} 
(c^\dagger_{jl\sigma}c^{}_{(j+1)l\sigma} + \mathrm{h.c.}) 
-t_\perp \sum_{\langle l l^\prime \rangle} \sum_{j\sigma}
c^\dagger_{jl\sigma}c^{}_{jl^\prime\sigma} \nonumber\\
& + U \sum_{jl} \hat{n}_{jl\uparrow} \hat{n}_{jl\downarrow} - \sum_{jl\sigma} \mu_\sigma \hat{n}_{jl\sigma}.
\end{align}
Here,
the index $j$ is used to label the lattice sites within a single 1D chain, while the chains
are labeled with the index $l$.
The operator $c^{}_{jl\sigma}$ ($c^\dagger_{jl\sigma}$) annihilates (creates) a fermion with
pseudo-spin $\sigma=\,\uparrow,\downarrow$ at site $j$ in chain $l$.
In the kinetic term, $t_\parallel$ and $t_\perp$ are the hoppings within the chain and
between the chains, respectively, while
the on-site interaction strength is denoted by $U$  and
the spin-dependent chemical potential by $\mu_\sigma$. 
In the following, 
we give all energies and temperatures in the units of $t_\parallel$, and set $t_\parallel = 1$.
At $t_\perp=0$ the system system is a collection of independent 1D chains whereas
at $t_\perp=1$ we have a 3D cubic lattice. The region $0<t_\perp<1$ then defines a
dimensional crossover from 1D to 3D. 
In this regime the system is infinite in all three spatial directions and
the emergence of long range order is possible.
We study the system with an attractive interaction $U=-3$.
Moreover, we describe the superfluid symmetry breaking using Nambu formalism.
Thus, in the following equations, the Green's function and
self-energy are interpreted in the form of $2\times 2$ Nambu blocks which are labeled
by the position.

We solve the equilibrium state of the system using a cluster variant of dynamical
mean-field theory (DMFT)~\cite{DMFTreview,clusterdmft}.  
In our cluster DMFT model we assume a periodic boundary condition within a single chain
\textit{and treat the whole chain as a single cluster in the algorithm}. 
In the directions perpendicular to the 1D chains, 
we assume that the self-energy of the system is local, i.e. 
\begin{align}
\bm{\Sigma}_{jj';ll'}(i\omega_n)=\delta_{l,l'}\bm{\Sigma}_{j j';l}(i\omega_n), 
\end{align}
where $i\omega_n$ is the Matsubara frequency.
In other words, we take the self-energy as block diagonal in the interchain index $l$.
This assumption reflects the fact that in the quasi-1D regime the dominant 
fluctuations occur in the intrachain direction. 
Notice also that this formulation is exact in the 1D limit. 
A similar approach has been utilized to study the Mott and Luttinger liquid transitions 
of the repulsive Hubbard model in quasi-1D lattices~\cite{chaindmft1,chaindmft2,chaindmft3}.
On the assumption that the system is homogeneous in the interchain direction, 
the self-energy is independent of the chain index $l$, and moreover,
the Green's function of the system is diagonal in the transverse quasi-momentum
$\bm{k}_\perp=(k_x,k_y)$.
The Dyson equation for the Green's function of the system is then given by
\begin{align}
&[\bm{G}(\bm{k}_\perp;i\omega_n)]^{-1}_{jj'} =\nonumber\\
&[\bm{G}^0_\parallel(i\omega_n)]^{-1}_{jj'}-\epsilon_{\bm{k}_\perp} \bm{\sigma}_z \delta_{jj'}  -\bm{\Sigma}_{jj'}(i\omega_n).
\label{eq:dyson}
\end{align}
Here, $\bm{G}^0_\parallel$ is the non-interacting Green's function of a single chain, 
while $\epsilon_{\bm{k}_\perp}$ is the transverse single particle dispersion given by
$\epsilon_{\bm{k}_\perp} \equiv -2t_\perp (\cos k_x + \cos k_y)$
and $\bm{\sigma}_z$ the Pauli $z$-matrix.
Taking a single chain as a cluster, 
the bath Green's function for the cluster DMFT becomes
\begin{align}
\left[\bm{\mathcal{G}}^0(i\omega_n)\right]^{-1}_{jj'}= 
\left[\sum_{\bm{k}_\perp} \bm{G}(\bm{k}_\perp;i\omega_n)\right]^{-1}_{jj'}+\bm{\Sigma}_{jj'}(i\omega_n).
\end{align}
We employ the continuous-time auxiliary-field quantum Monte Carlo method in Nambu
formalism to solve the
impurity problem of the DMFT iteration~\cite{Gull2008,Gull2011,Koga2010}.
That is, within the cluster, all local and non-local fluctuations are taken into account.
To facilitate the large expansion order imposed
by the low temperature and the large cluster size of the simulations, we utilize 
delayed spin-flip update~\cite{Alvarez2008} and submatrix update~\cite{submatrix} 
techniques to speed up the computation.

Our approach allows us to study the superfluid pairing on a general basis, 
including also the possibility of spatially non-uniform solutions. 
For example, in the presence of spin-polarization the two-component
Fermi gas may enter the Fulde-Ferrell-Larkin-Ovchinnikov
(FFLO) phase which involves spontaneous breaking of the translation invariance 
of the superfluid state~\cite{FF,LO}. 
To give a point of contrast to alternative approaches, 
the cellular DMFT method could be criticized here for the
explicit breaking of translation invariance on the level of the method, which might favor
states with broken translation invariance. On the other hand, enforcing the spatial symmetry as in 
the dynamical cluster approximation (DCA) would contain precisely the opposite problem. 
Thus, we chose to adopt a periodic boundary condition which
allows for solutions with broken translation invariance without introducing any such
broken symmetry on the level of the computational method.

We define the superfluid order parameter as 
$\Delta_j=-U\langle c^\dagger_{j,\uparrow} c^\dagger_{j,\downarrow}\rangle$.
Here, we identify the BCS state as the state with non-zero and uniform $\Delta_j$
over the whole system. Since $\Delta_j$ is defined as an anomalous expected value,
this criterion also implies long range order. The FFLO state is defined as the state
with $\Delta_j$ oscillating with position. In order to study the FFLO mechanism
we vary the spin-polarization $P=(N_\uparrow-N_\downarrow)/(N_\uparrow+N_\downarrow)$
through the spin-dependent chemical potentials 
while keeping the total particle number constant at half-filling.

The phase diagram of the system is presented in Fig.~\ref{fig:phasediag}(b).
At $t_\perp=0.3$ the BCS critical temperature obtained from cluster DMFT
is $T_c=0.12$ and decreases monotonously 
as the interchain hopping is reduced reaching a value of $T_c=0.05$ at $t_\perp=0.1$.
Below $t_\perp=0.1$, we are limited by the computational cost of 
the impurity problem at cluster sizes and temperatures relevant for 
the BCS transition.
However, the results for finite $t_\perp$ suggest convergence to a critical temperature
of zero at the 1D limit, as is expected because of the Mermin-Wagner
theorem.

To quantify the effect of the non-local quantum fluctuations on the BCS state, 
let us compare the result to the single-site DMFT calculations.
We compute the phase diagram
of the system using single-site real-space DMFT~\cite{Helmes2008,Snoek2008,Koga2011rdmft,Kim2011}, 
where the main assumption is that the self-energy is local. 
In this approximation, the expression for the self energy above simplifies further to
$\bm{\Sigma}_{jj'}(i\omega_n)=\delta_{j,j'}\bm{\Sigma}_{j}(i\omega_n)$.
Note that the self-energy is still frequency dependent,
i.e.~the model contains all local quantum fluctuations.
The model reduces to static mean-field theory if also the $i\omega_n$ dependence is completely discarded.
The lattice Dyson equation remains in the same form as given in Eq.~\eqref{eq:dyson}, 
while the quantum impurity model of DMFT
is now reduced to a single site problem with a bath Green's function given by
\begin{align}
[\bm{\mathcal{G}}^0_{j}(i\omega_n)]^{-1} = 
\left[\sum_{\bm{k}_\perp} \bm{G}_{jj}(\bm{k}_\perp;i\omega_n)\right]^{-1}+\bm{\Sigma}_{j}(i\omega_n).
\end{align}
Notice that each matrix element above is still a $2\times 2$ Nambu block 
of the normal and anomalous on-site Green's functions or self-energies. 
Again, the reason for using the real-space formulation is that it allows
us to describe also superfluid states with spatial symmetry breaking.

The single-site DMFT predicts a nearly constant critical temperature for the BCS state with
$T_{c}\approx 0.14$ over the corresponding parameter range. It is then readily apparent
that the rapid disappearance of superfluidity in the cluster model cannot be
attributed to changes in the non-interacting density of states caused by the 
varying dimensionality. Such effects would already be included to the single-site model.
Here, it should be noted that already the local quantum fluctuations bring a substantial
correction the static mean-field prediction of the critical temperature $T_{c,\textrm{MF}}=0.52$,
though qualitatively the static mean-field and single-site DMFT critical temperatures 
behave similarly as a function of the interchain hopping.

\begin{figure}
\includegraphics[width=0.4\textwidth]{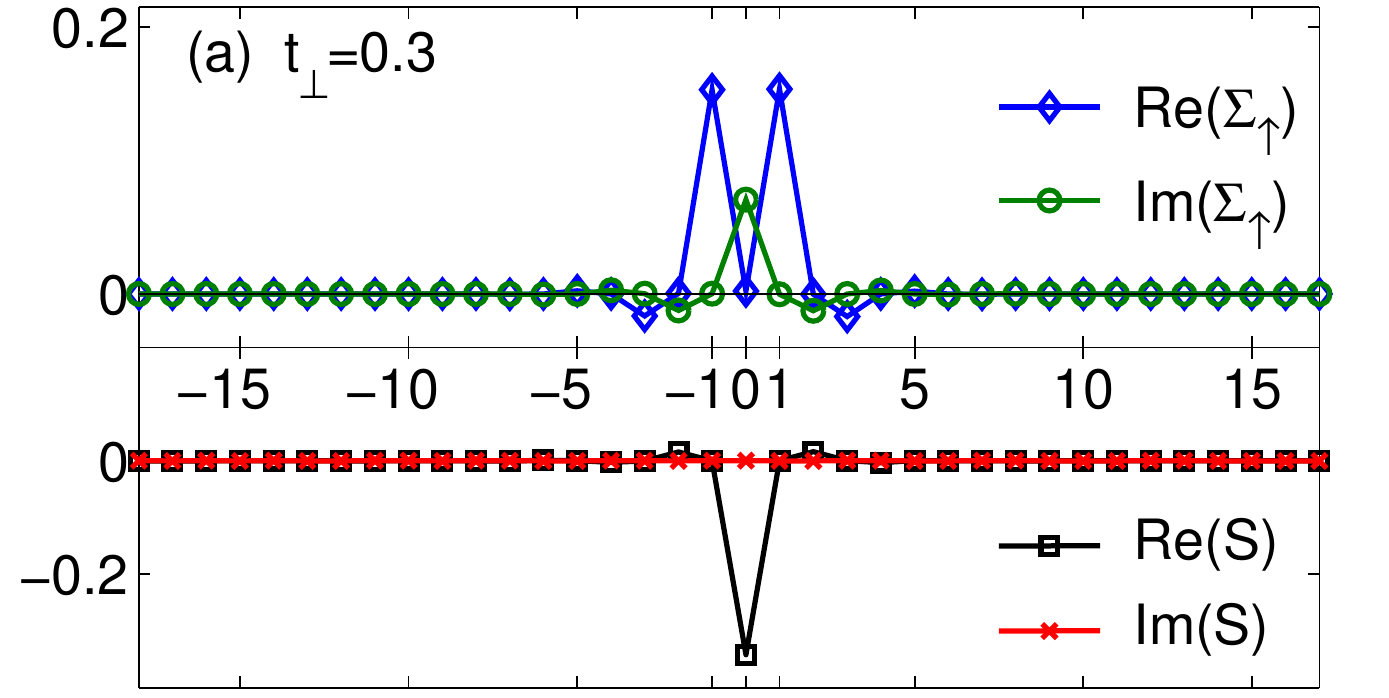}
\includegraphics[width=0.4\textwidth]{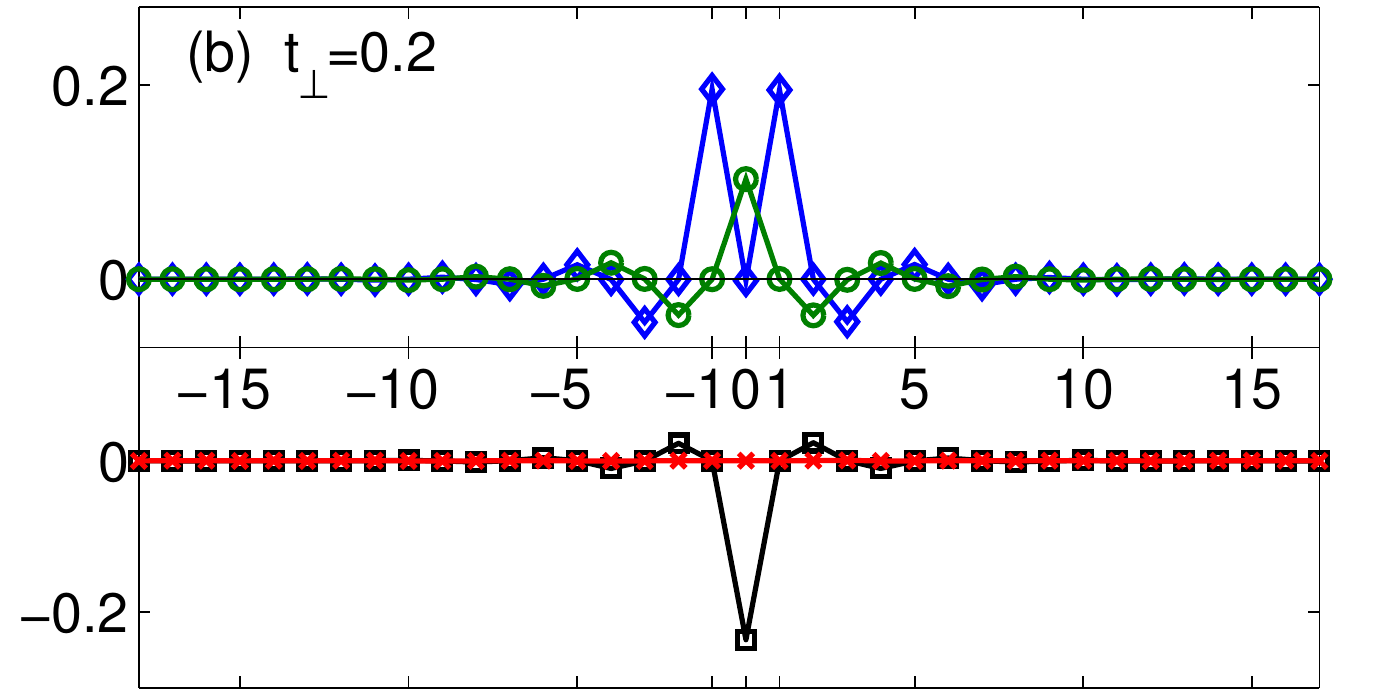}
\includegraphics[width=0.4\textwidth]{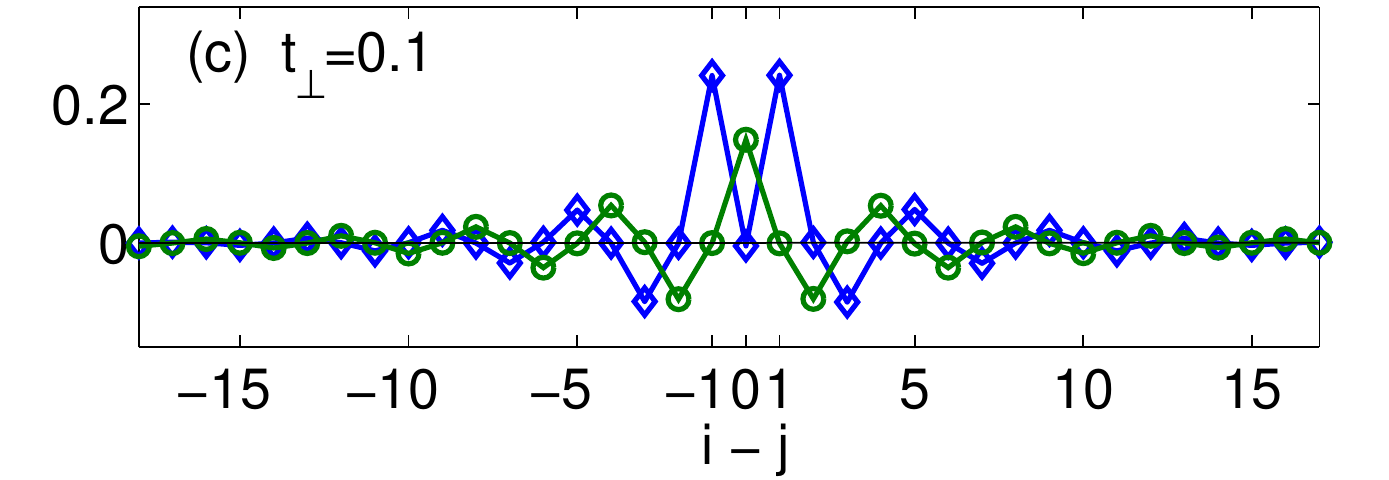}
\caption{\label{fig:nonloc}
The dependence of the self-energy on the dimensionality.
In each panel we plot the normal spin-$\uparrow$ and anomalous Nambu components,
$\Sigma_{\uparrow,ij}(i\omega_n)$ and $S_{ij}(i\omega_n)$, of the self-energy
between sites $i$ and $j$ for the lowest Matsubara frequency.
Here, the system is at a temperature
of $T=0.08$ and at half-filling with zero spin-polarization. At (a) $t_\perp=0.3$ 
and (b) $t_\perp=0.2$ this corresponds to the BCS state, while at (c) $t_\perp=0.1$
the system is in the normal phase 
in which case the anomalous self-energy is identically zero and thus not plotted.
The non-local quantum fluctuations
grow substantially as the interchain hopping $t_\perp$ decreases.
The oscillating structure of the self-energy in $i-j$ is
predominantly a single particle signature
and corresponds to the Fermi momentum of the system. 
Note that the BCS order parameter is chosen real and 
therefore the BCS anomalous self-energy is also real.
}
\end{figure}

In Fig.~\ref{fig:nonloc} we plot the self-energy of the system 
at a constant temperature while varying the interchain hopping $t_\perp$.
The figure demonstrates that the non-local component of the self-energy grows rapidly towards
the 1D limit. Therefore, we may conclude that
the drastic decline of the superfluid critical temperature
is driven by the non-local quantum fluctuations.
Moreover, Fig.~\ref{fig:nonloc} indicates that the cluster size of our simulations is sufficient
to exhaust the self-energy of an individual chain.

Let us now turn to the case of spin-polarized systems.
The fact that quasi-1D systems would favor the FFLO state in comparison to 3D systems was
first suggested based on mean-field studies of a system of coupled 1D tubes~\cite{parish2007}.
The same qualitative conclusion was reached in~\cite{zhao2008} using effective field theory
and treating the intertube coupling as a perturbation. On the other hand, real-space DMFT studies of
coupled chains~\cite{Kim2012,Heikkinen2013} suggested rather that the FFLO state
is important in the entire dimensional crossover from quasi-1D to 3D lattices. 
One reason for differing predictions can be that the stabilization of FFLO
in lattices due to nesting~\cite{Koponen2007} is stronger for coupled chains than
coupled tubes. Another possibility is a different treatment of quantum fluctuations.
While the FFLO signatures have been absent in
experiments on spin-polarized Fermi gases in continuum~\cite{Zwierlein2006,Partridge2006},
experiments in 1D tubes~\cite{Liao2010} are consistent with its possible existence.

Now, considering the large effect of the quantum fluctuations on the BCS transition,
one might anticipate that the FFLO state which involves a delicate spatial symmetry
breaking would be totally destroyed by the non-local quantum fluctuations.
Here we show that, in fact, the FFLO state survives even in the presence of non-local
quantum fluctuations, as shown in Fig.~\ref{fig:phasediag}(b). The qualitative
trend is similar to the BCS transition; the critical temperature of the FFLO
phase decreases when the interchain hopping is reduced. 
At $t_\perp=0.2$ we find that the critical temperature
of FFLO in the cluster model is lowered by a factor of $0.67$ in comparison
to the single-site approximation, while for the BCS critical temperature
the corresponding ratio would be $0.69$.
At the critical temperatures reported in Fig.~\ref{fig:phasediag}(b)
the polarization of the system varies from $P=3~\%$ at $t_\perp=0.2$
to $P=4~\%$ at $t_\perp=0.3$ in the cluster simulations, whereas in the
case of single-site DMFT we find a polarization of $P=6~\%$ in the same range. 
Below $t_\perp=0.2$ we cannot reach the FFLO phase in our simulations as we 
are limited by the scaling of the computational cost.

\begin{figure}[h!]
\includegraphics[width=0.44\textwidth]{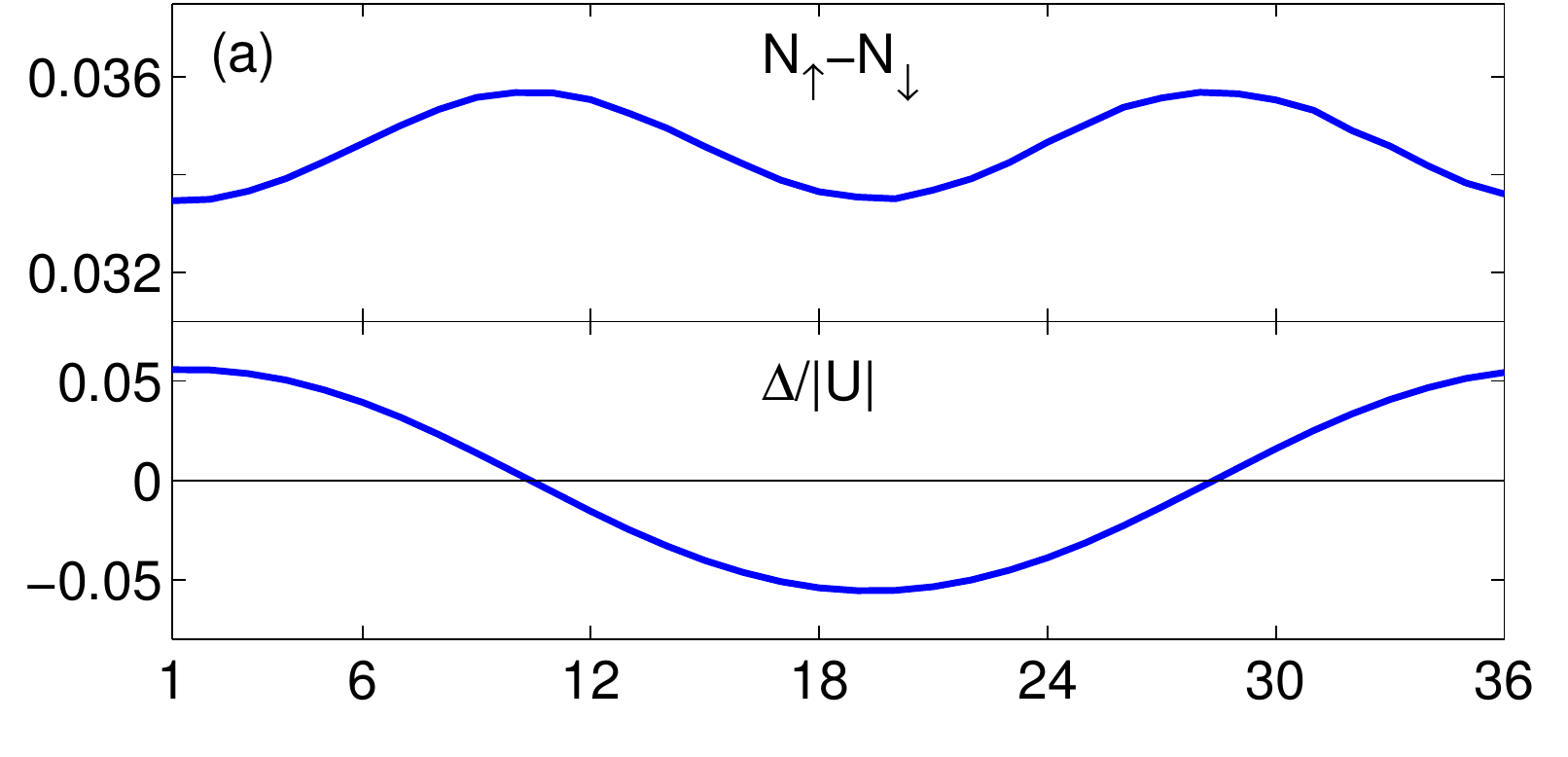}
\includegraphics[width=0.44\textwidth]{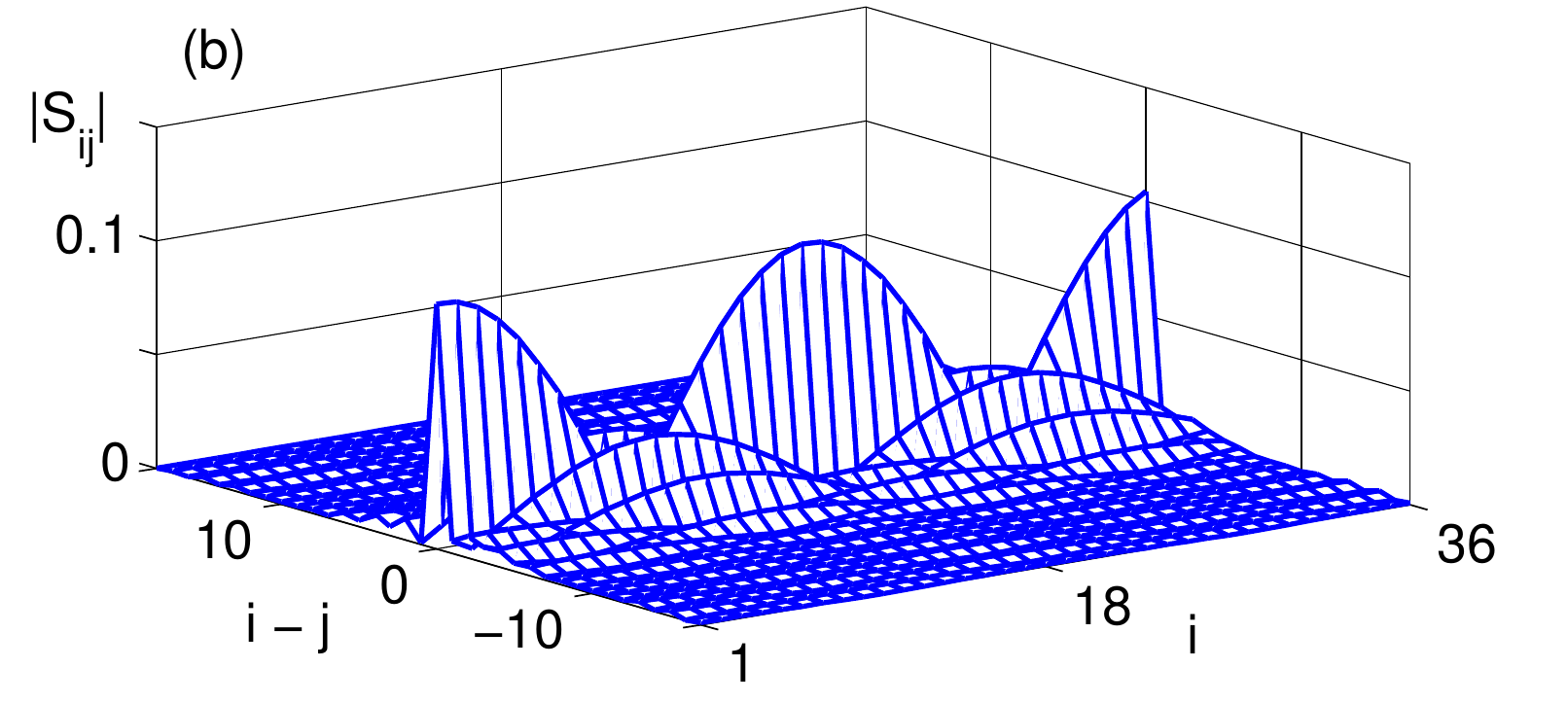}
\caption{\label{fig:fflophysics}
The FFLO state at $t_\perp=0.25$. Here, the
temperature is $T=0.05$ and the polarization $P=0.035$.
(a) The density difference $N_\uparrow-N_\downarrow$ and the 
order parameter $\Delta$ as a function of the cluster site $i$.
In our results, the translation invariance of the system is spontaneously broken and
in a particular simulation in the FFLO regime e.g. the minima of the order parameter
and density may fall on any given lattice site.
(b) The absolute value of the anomalous part of the 
self-energy, $S_{ij}(i\omega_n)$,
plotted as a function of the cluster site $i$ and the distance $i-j$ 
to the site $j$ for the lowest Matsubara frequency. 
}
\end{figure}

Throughout the data, we find $\Delta_j$ in FFLO state
an approximately sinusoidal function. Moreover, we find that the oscillating
order parameter is accompanied by a spatial modulation of the density with half
the period of the order parameter, as demonstrated in Fig.~\ref{fig:fflophysics}(a).
This agrees with the standard characterization of the FFLO state.
In Fig.~\ref{fig:fflophysics}(b) we the structure of the non-local
part of the self-energy in the FFLO state.
The similar, approximately sinusoidal, dependence on the position is
found at all Matsubara frequencies, while the contribution of the non-local
fluctuations is the largest at low frequencies, as expected from the
analytical high frequency asymptotes. The fact that the cluster self-energy retains
the same periodic structure at all frequencies, and furthermore, that the local part
of the self-energy is dominant is suggesting that FFLO character of the 
many-body state is robust and experimentally discernible from a polarized
superfluid by probes or imaging techniques sensitive to density modulations.

There is an additional point to be made about the convergence with cluster size
in the spin-polarized case. In our simulations,
we do not find the FFLO phase at cluster sizes below $N_c\approx 30$. The likely reason here
is that the lowest possible non-zero pairing momentum, $\bm{q}=2\pi/N_c$, leads to a too large
increase in the kinetic energy of the Cooper pairs at $N_c \lesssim 30$. In other words, the 
FFLO transition is in essence orbitally limited in small clusters. 
We also investigated cluster sizes larger than $N_c=36$ up to $N_c=42$ at $t_\perp=0.3$ and found 
no change in the FFLO transition suggesting that the cluster size $N_c=36$ is sufficient.
Finally, it is interesting to speculate, if the non-local quantum fluctuations can in fact be favorable
for the spatial symmetry breaking of the FFLO state by destabilizing the spatially uniform BCS state. 
The data at $t_\perp=0.3$ of Fig.~\ref{fig:phasediag}(b) does in fact suggest a scenario along these lines. 
However, drawing this conclusion fully would require a further analysis 
of the role of the interchain quantum fluctuations on the FFLO state
by extending the cluster formalism beyond the single chain approximation,
and is beyond the scope of the present work.

In summary, we have shown that non-local quantum fluctuations play a
crucial role in the low temperature properties of the attractive Hubbard model 
and affect heavily the fermionic superfluidity.
Still, we find that even the exotic FFLO superfluid with broken translation invariance 
can endure the effect of the fluctuations, and possibly even compete better with the uniform
polarized superfluid state because of the fluctuations.
Our results suggest that the build-up of non-local quantum fluctuations can be studied in a systematic
way in experiments on ultracold atoms in anisotropic optical lattices.
An interesting future direction would also be to study the interplay of non-local quantum fluctuations
and nearest neighbour interactions. The experimental study of such interactions is evolving rapidly 
at the moment~\cite{Baranov2012review}, and they are likely to have important implications 
to the phase diagram of the system based on 1D predictions~\cite{Aligia2000}.

\begin{acknowledgments}
This work was supported by the Academy of Finland through 
its Centers of Excellence Programme (2012-2017) 
and under Projects No. 139514, No. 141039, 
No. 263347, No. 251748, No. 272490 and No. 135000,
and by the European Research Council (ERC-2013-AdG-340748-CODE
and ERC-2011-AdG 290464-SIMCOFE).
M.O.J.H. acknowledges support from the KAUTE-foundation.
This research is supported in part by GIST college's 2013-2014 GUP 
research fund.
Computing resources were provided by CSC--the Finnish IT 
Centre for Science and the Triton cluster at Aalto University. 
\end{acknowledgments}

%\bibliography{finitetemp_fflo_bibfile}

\end{document}